# Design Optimization of eVTOL Propellers using a Viscous-Extension Discrete Vortex Method

**Rahul Kumar** [a,1] and **Ramkumar Pathmanabhan** [a,2]

[a] *Ubifly Technologies Private Limited, Chennai, Tamilnadu-600113, India*

*Corresponding author:

[1] E-mail: rahul.kumar@eplane.ai

[2] E-mail: ramkumar@eplane.ai

**Abstract**

Potential flow theory remains a cornerstone of unsteady aerodynamics due to its computational efficiency in modeling complex flow phenomena. This study presents a significant advancement by integrating a viscous unsteady theory with established numerical vortex methods, creating a hybrid computational tool for low-to-moderate Reynolds number flows. We develop a Viscous Discrete Vortex Method (VDVM) by replacing the classical inviscid Kutta condition with a closure derived from triple-deck boundary layer theory, allowing the model to account for Reynolds number dependencies and unsteady viscous effects. The framework utilizes a three-dimensional vortex ring scheme and an unsteady Bernoulli formulation for load calculation. The model is validated against experimental and high-fidelity CFD data, showing excellent agreement in thrust and torque across a wide operational envelope. Using this validated framework, we conduct a systematic parametric investigation into rotor blade design for electric vertical take-off and landing (eVTOL). A sophisticated optimization of the spanwise geometry was performed: twist distributions were calculated by iteratively solving for axial and tangential induction factors to maintain optimal local angles of attack, while chord distributions were derived using the Adkins and Liebeck framework to satisfy the Betz condition for maximum efficiency. Results demonstrate that this tapered chord and nonlinear twist profile significantly mitigate tip losses and manage spanwise loading. The optimized geometry achieved an 8.99% increase in the efficiency compared to the baseline. This work bridges the gap between high-fidelity viscous analysis and fast vortex methods, providing a versatile tool for the performance-driven design of lifting surfaces in unsteady flight regimes.

## 1. Introduction

Potential flow theory remains a foundational pillar of aeronautical engineering, evolving from the early insights of d'Alembert and Euler [1] to the definitive works of Prandtl [2]. This theory allows engineers to model a wide variety of aerodynamic problems by using analytical and numerical methods to simplify complex fluid behaviors. Traditional models such as thin airfoil theory [3], lifting-line theory [4], and the Wagner [5] and Theodorsen formulations [6] frequently result in complex singular integral equations. To solve these efficiently, researchers like Belotserkovskii [7] pioneered discretization techniques that convert these integrals into manageable systems of linear algebraic equations [8]. This mathematical shift is the core of the Vortex Lattice Method (VLM), a highly popular and computationally lean tool for steady aerodynamic analysis [9]. In VLM, an aircraft's lifting surface is divided into a grid of panels, each containing a "bound vortex." By enforcing a no-penetration boundary condition (ensuring air doesn't flow through the wing), the strength of these vortices can be calculated. For dynamic scenarios involving pitching or surging motions, the Unsteady Vortex Lattice Method (UVLM) [10], [11], [12], [13] builds on this by accounting for the wake. It continuously sheds vortices from the trailing edge and tracks their movement over time to maintain an accurate physical representation of the airflow. Ultimately, these methods balance mathematical rigor with practical efficiency, remaining essential for both classical study and modern aerodynamic design.

An alternative vortex-based approach for modeling unsteady aerodynamics is the discrete vortex method (DVM) [14], [15], [16], [17], [18]. In contrast to the UVLM, the DVM determines bound circulation directly, thereby circumventing the solution of a linear system at each time step. This method relies on a conformal mapping between the airfoil and a circular cylinder, leveraging an analytical potential-flow solution. The primary unknown at each instant is the strength of the vortex shed from the trailing edge, typically prescribed by enforcing the Kutta condition. While this approach eliminates matrix inversion, it is inherently limited to geometries for which an explicit conformal mapping exists, precluding its application to arbitrary or dynamically deforming shapes. Moreover, DVM, are widely used in research and engineering practice due to their capacity to capture dominant flow features at a low computational cost. While traditionally based on inviscid and attached-flow assumptions, significant efforts have been made to extend these methods to more complex regimes such as high angles of attack and flow separation [19], [20]. These conditions are frequently encountered in modern applications, including bio-inspired flight [10], [11], [12], and dynamic stall [21], [22].

The classical Kutta condition, while a staple for steady attached flows at high Reynolds numbers, frequently fails to capture the physical realities of highly unsteady or low-Reynolds-number regimes has been questioned in numerous studies [23], [24], [25]. Recent research [26], [27] addresses these shortcomings by developing a viscous extension to potential flow theory, specifically by relaxing the rigid requirement that the trailing-edge singularity vanish. Instead, they utilize triple-deck boundary layer theory to determine the singularity's strength based on the Reynolds number and a viscous unsteady effective angle of attack. This paper builds upon that semi-analytical "viscous theory" which was originally limited to flat plates by integrating it into numerical vortex methods like UVLM and DVM. By replacing the standard Kutta closure with this triple-deck formulation, the new model enables the analysis of arbitrary body shapes and wake deformations. This integrated approach is validated against classical benchmarks, such as the Wagner [5] and Theodorsen [6] formulations, providing a more versatile tool for modern aeronautical engineering applications.

This paper develops and validates a Viscous-Extension Discrete Vortex Method (VDVM) for low-to-moderate Reynolds number rotor flows. While viscous-extension methods have seen success in harmonic-motion applications, this work represents a significant leap by adapting the VDVM architecture to the complex, multi-axial flow fields of rotating geometries. We begin by detailing the integration of a viscous unsteady theory with a three-dimensional vortex ring scheme, where the traditional inviscid Kutta condition is replaced by a closure derived from triple-deck boundary layer theory. Following the description of the model and the unsteady Bernoulli formulation for load calculation, the framework is validated against both experimental data and high-fidelity CFD results. The paper then transitions into a parametric investigation of rotor blade geometry. Specifically, we describe an optimization process that employs a nonlinear twist distribution to maintain optimal local angles of attack and an Adkins and Liebeck-based chord distribution to satisfy the Betz condition [28]. We believe that this VDVM framework offers a versatile tool for the performance-driven design of lifting surfaces in the unsteady flight regimes typical of eVTOL and UAV platforms, particularly where high-fidelity viscous analysis must be balanced with the speed required for large-scale shape optimization.

**2. NUMERICAL METHODOLOGY: DISCRETE VORTEX METHOD**

In this analysis, a three-dimensional rotor blade simulation is carried out using the Viscous discrete vortex method (VDVM), which is well suited for accurately modeling unsteady potential flow [29], [30], [31], [32], [33]. In this approach, the rotor blade is discretized into a large number of quadrilateral flat surface panels, as shown in Fig. 1. Specifically, the rotor blade is divided into an $N \times M$ grid, where $N$ panels are distributed along the spanwise direction and $M$ panels along the chordwise direction. The bound circulation on the rotor blade and the associated vortex wake are represented using vortex ring elements. For each panel, the leading segment of the vortex ring is positioned at the quarter-chord location (Fig. 1), while the collocation point, at which the no-penetration boundary condition is enforced, is located at the three-quarter-chord position of the panel. The normal vector $n$ is defined at the collocation point, which also serves as the center of the vortex ring for each panel. At any spatial point ($x$, $y$, $z$), the induced velocity components ($u$, $v$, $w$) are computed based on the vortex ring formulation. Initially, vortex rings are present only on the rotor blade panels, and during the first time step, the trailing vortex segment at the trailing edge acts as the starting vortex due to the absence of free wake elements. As the simulation advances, the trailing edge of the rotor blade moves, and new wake vortex rings are formed by connecting the updated aft points of the trailing-edge vortex rings. This wake-shedding process generates a new set of trailing-edge wake vortex rings at each time step. By applying Kelvin's circulation theorem, each trailing-edge panel sheds a wake panel with vortex strength equal to its circulation from the previous time step.

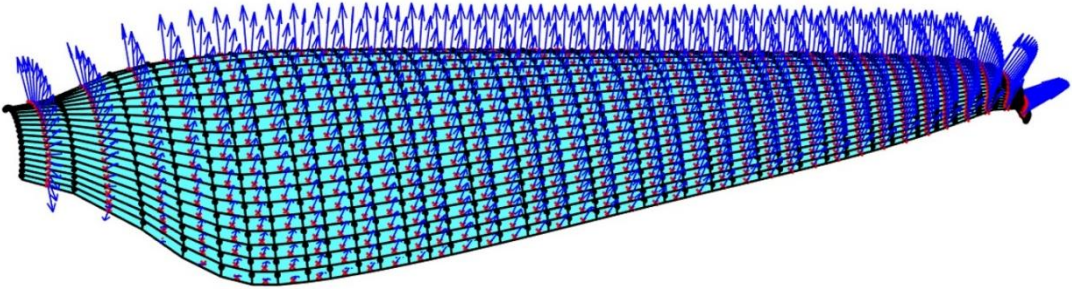

Fig. 1. Illustration of vortex and collocation point. At each quarter of the panel leading segment of the vortex ring is located. The collocation points where no normal flow condition is applied is located at the center of the three fourth of the panel.

**2.1 Vortex Dynamics and Formulation**

The three-dimensional, incompressible, and inviscid flow field is governed by the vorticity transport equation:

$$\frac{\partial \Omega}{\partial t} + \mathrm{u} \cdot \nabla \Omega = \Omega \cdot \nabla \mathrm{u} + \nu \nabla^2 \Omega, \quad (1)$$

where $\mathrm{u}(\mathrm{x}, t)$ is the velocity field and $\Omega(\mathrm{x}, t) = \nabla \times \mathrm{u}$ is the vorticity field. Under the inviscid potential flow assumption, the viscous diffusion term ($\nu \nabla^2 \Omega$) is neglected. For three-dimensional flows, the vortex stretching and tilting term ($\Omega \cdot \nabla u$) is retained, leading to the Lagrangian vorticity evolution equation:

$$\frac{D\Omega}{Dt} = \Omega \cdot \nabla u, \quad (2)$$

A Lagrangian approach is employed to track the evolution of the wake vortices shed from the lifting surface.

The velocity induced ($\vec{dv}$) by a finite vortex segment of circulation $\Gamma$ and length $dl$ at a point located a distance r away is given by the Biot-Savart law:

$$\vec{dv} = \frac{\Gamma}{4\pi}\left[\frac{\vec{dl} \times \vec{r}}{|\vec{r}|^3}\right], \quad (3)$$

To avoid singularities when vortex elements approach each other, a finite-core vortex model or a regularization scheme is typically incorporated.

The influence coefficient $a_{KL}$ quantifies the normal velocity induced at collocation point $K$ by a vortex ring $L$ of unit strength:

$$a_{KL} = (u, v, w)_{KL} \cdot n_K, \quad (4)$$

where $(u, v, w)_{KL}$ are the velocity components induced at point $K$ by the vortex ring $L$. Enforcing zero normal flow through the surface at all collocation points yields the linear system:

$$\sum_{L=1}^{m} a_{KL}\, \Gamma_L = \text{RHS}_K, \text{ for } K = 1,2,\ldots,m, \quad (5)$$

where $m = M \times N$. The right-hand side term $\text{RHS}_K$ accounts for the normal component of the freestream velocity and the velocity induced by the entire wake structure at collocation point $K$:

$$\text{RHS}_K = -[U(t) + u_w]_K \cdot n_K, \quad (6)$$

Here, $U(t)$ is the kinematic velocity of the wing due to its motion, and $u_w$ is the velocity induced by all wake vortex elements. The linear system is assembled in matrix form as:

$$A\Gamma = b, \quad (7)$$

and solved at each time step to obtain the unknown bound vortex strengths $\Gamma = \{\Gamma_1, \Gamma_2, \ldots, \Gamma_m\}^T$.

## 2.2 Calculation of Aerodynamic Loads

Once the bound circulation distribution is determined, the pressure jump across each panel is computed using the unsteady Bernoulli equation. For a vortex-lattice discretization, the pressure difference $\Delta p_{ij}$ for panel $(i, j)$ is given by:

$$\Delta p_{ij} = \rho \left\{ [U(t)+u_w]_{ij} \cdot \tau_i \frac{\Gamma_{i,j} - \Gamma_{i-1,j}}{\Delta c_{ij}} + [U(t)+u_w]_{ij} \cdot \tau_j \frac{\Gamma_{i,j} - \Gamma_{i,j-1}}{\Delta b_{ij}} + \frac{\partial}{\partial t}\Gamma_{i,j} \right\}, \quad (8)$$

where $\rho$ is the fluid density, $\Delta c_{ij}$ and $\Delta b_{ij}$ are the panel chord and span, and $\boldsymbol{\tau}_i, \boldsymbol{\tau}_j$ are the panel tangential vectors in the chordwise and spanwise directions. The terms represent chordwise and spanwise contributions of the tangential velocity, and the time rate of change of circulation, which is critical for capturing unsteady effects.

The total thrust is obtained by integrating the pressure forces over all panels:

$$T = \sum_{j=1}^{N}\sum_{i=1}^{M} \Delta p_{ij} S_{ij} \cos \alpha_{ij}, D$$

$$= \sum_{j=1}^{N}\sum_{i=1}^{M} \rho \left[ (w_{ind} + w_w)_{ij}(\Gamma_{i,j} - \Gamma_{i-1,j})\Delta b_{ij} + \frac{\partial \Gamma_{ij}}{\partial t}\Delta S_{ij}\sin \alpha_{ij} \right], \quad (9)$$

The total torque over all panels is defined as:

$$Q = \sum_{j=1}^{N}\sum_{i=1}^{M} (\Delta p_{ij} S_{ij} \sin \alpha_{ij} \cdot r_j)$$

$$+ \sum_{j=1}^{N}\sum_{i=1}^{M} \rho \left( (w_{ind} + w_w)_{ij}(\Gamma_{i,j} - \Gamma_{i-1,j})\Delta b_{ij} + \frac{\partial \Gamma_{ij}}{\partial t}\Delta S_{ij} \cos \alpha_{ij} \right)$$

$$\cdot r_j, \quad (10)$$

The total aerodynamic power over all panels is defined as:

$$P = 2\pi \cdot n \cdot Q, \quad (11)$$

where, $n$ is the rotational speed (rev/s), $S_{ij}$ is the panel area, $\alpha_{ij}$ is the local angle of attack, and $w_{ind}$ is the induced downwash.

Table 1: The non-dimensional coefficients used throughout this work are defined as follows:

| Symbol | Definition | Description |
|---|---|---|
| $J$ | $U_\infty / (nD)$ | Advance ratio |
| $C_T$ | $T / (\rho n^2 D^4)$ | Thrust coefficient |
| $C_Q$ | $Q / (\rho n^2 D^5)$ | Torque coefficient |
| $C_P$ | $P / (\rho n^3 D^5) = 2\pi C_Q$ | Power coefficient |
| $\eta$ | $J \cdot C_T / C_P$ | Propulsive efficiency |

## 3. VISCOUS EXTENSION OF THE DISCRETE VORTEX METHOD

### 3.1 Viscous Correction to the Classical Kutta Condition

The classical Kutta condition, which enforces smooth flow at a sharp trailing edge by eliminating the velocity singularity, is a cornerstone of potential flow theory. However, its validity is questionable in flows characterized by viscous-inviscid interaction [23], [24], [25]. To overcome this limitation, the present work integrates a rigorous viscous correction derived from triple-deck boundary layer theory, as developed by Taha and Rezaei [27].

This model relaxes the classical Kutta condition by allowing a finite, physics-determined strength of the trailing-edge singularity. The strength, denoted $B_v(t)$, is not set to zero but is calculated from a matched asymptotic analysis that resolves the interaction between the Blasius boundary layer on the body and the nascent shear layer in the wake. This introduces an explicit dependence on the Reynolds number $Re$ and an effective angle of attack $\alpha(t)$.

For a surface, the tangential velocity distribution with a relaxed Kutta condition is:

$$\frac{u(x,t)}{U} = \frac{1}{\sqrt{1-(x/b)^2}} [\alpha_s(t) \mp B_v(t)], \quad (12)$$

where the $\mp$ sign applies to the suction and pressure sides, respectively. The term $\alpha_s(t)$ is the effective quasi-steady angle of attack from the instantaneous potential flow solution.

The viscous correction parameter is given by:

$$B_v(t) = -2\varepsilon^3 \kappa^{-\frac{5}{4}} \left[ \frac{1}{2} a_0(t) + \sum_{n=1}^{\infty} n\, a_n(t) \right] \mathcal{B}_e(\alpha), \quad (13)$$

Here, $\varepsilon = Re^{-1/8}$, $\kappa \approx 0.334$ is the Blasius skin-friction coefficient, and $a_n(t)$ are the Fourier coefficients of the inviscid pressure distribution. The nonlinear function $\mathcal{B}_e(\alpha_e)$, obtained from the numerical solution of the triple-deck equations [27], encapsulates the viscous response and features a stall asymptote at a critical effective angle.

### 3.2 Implementation of the Viscous Closure in the DVM

Within the framework of the DVM, the viscous correction manifests as an additional vortex, $\Gamma_v = 2\pi b U B_v(t)$, placed at the center of the circle in the conformal mapping plane that transforms the cylinder to the actual body geometry. This modifies the fundamental equation for conservation of circulation.

In the classical DVM, the strength of the nascent wake vortex $\Gamma_1$ is determined by enforcing the Kutta condition, leading to a relation where the total bound quasi-steady circulation $\Gamma_0$ is balanced: $\Gamma_k = -\Gamma_0$. With the viscous extension, this balance is altered to account for the viscous vortex $\Gamma_v$:

$$\Gamma_k = -\Gamma_0 - \Gamma_v \quad (14)$$

Consequently, the strength of the vortex shed from the trailing edge at each time step is computed from the modified relation:

$$\Gamma_1 = -\text{Im}\left[\frac{r}{\zeta_0 - r}(\Gamma_0 + \Gamma_v) + r\sum_{n=2}^{N_w}\Gamma_n\left(\frac{1}{\zeta_0 - \zeta_n} - \frac{1}{\zeta_0 - \frac{r^2}{\overline{\zeta_n}}}\right)\right] \quad (15)$$

where Im denotes the imaginary part of a complex number, $\zeta_0$ is the nascent vortex position in the circle plane, $r$ is the circle radius, and $\Gamma_n$ are the strengths of previously shed wake vortices at positions $\zeta_n$.

This modification ensures that the shed circulation is consistent with both the instantaneous kinematics of the body and the viscous state of the trailing-edge flow. The viscous correction $\Gamma_v(t)$ is updated at each time step based on the current potential flow solution (which yields $a_0(t)$ and $a_n(t)$) and the Reynolds number.

## 4. Methodology Validation

The numerical model was validated by comparing experimental data obtained through wind tunnel testing at the Indian Institute of Science, Bengaluru with results from the Viscous Discrete Vortex Method (VDVM). The experimental setup, featuring a precision CNC-machined aluminium propeller of diameter 0.30 m was mounted on a rigid sting balance within an open-circuit wind tunnel, is shown in Fig. 2, driven by a Neu-8055 motor and controlled via an Arduino Mega 2560. To characterize static performance, the thrust and torque were measured across a rotational speed range of 1330–8841 RPM; these results are presented in Table 2. For each throttle setting, the system was allowed to stabilize for two seconds before sampling load cell data at 1 Hz over a 5-second duration, while electrical parameters and RPM were simultaneously captured from the motor controller's buffer. The VDVM predictions demonstrate excellent alignment with experimental data. $C_T$ deviations remain minimal, ranging between 1.99% and 4.3%, while $C_Q$ values are consistently overpredicted by 7.95% to 14.80%. This discrepancy in torque is likely due to inherent measurement sensitivity and specific modeling assumptions. Ultimately, the close agreement across the operating range, particularly at higher RPMs, confirms the reliability of the VDVM for predicting propeller performance and justifies its application in subsequent aerodynamic design studies.

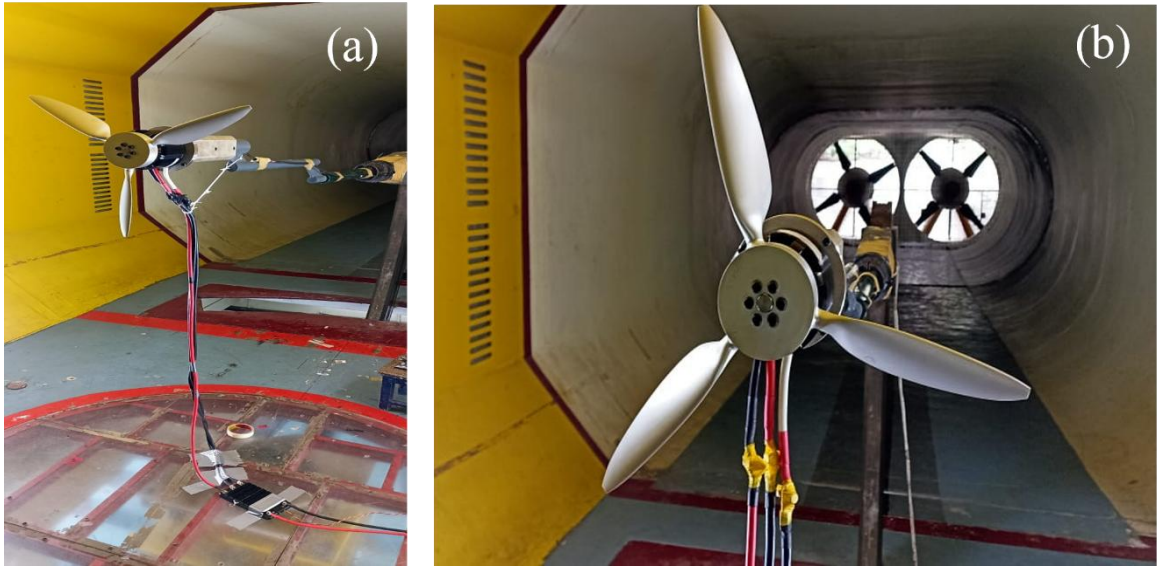

Fig. 2 Propeller-motor assembly mounting within the wind tunnel test section.

Table 2: Comparison VDVM with experimental results

| $n$ | $C_T$ (EXP) | $C_T$ (VDVM) | $C_Q$ (EXP) | $C_Q$ (VDVM) | $\Delta C_T\%$ | $\Delta C_Q\%$ |
|---|---|---|---|---|---|---|
| 1330 | 0.13958 | 0.14563 | 0.02977 | 0.03213 | 4.33 | 7.95 |
| 2126 | 0.16663 | 0.17338 | 0.02907 | 0.03238 | 4.05 | 11.37 |
| 3580 | 0.18177 | 0.18882 | 0.02795 | 0.03208 | 3.88 | 14.80 |
| 4921 | 0.18881 | 0.19477 | 0.02727 | 0.03096 | 3.16 | 13.54 |
| 6353 | 0.19091 | 0.19686 | 0.02686 | 0.03056 | 3.12 | 13.81 |
| 7682 | 0.19863 | 0.20227 | 0.02701 | 0.03053 | 1.83 | 13.06 |
| 8841 | 0.20025 | 0.20424 | 0.02720 | 0.03064 | 1.99 | 12.62 |

The VDVM results were further validated by comparing them with CFD predictions across various rotational speeds ($n$ =1500 to 3200 rpm) and advance ratios ranging from 0 to 0.91, summarized in Table 3. For all tested conditions, the VDVM demonstrates a high level of consistency in predicting $C_T$, with deviations from CFD results remaining remarkably low, generally between 3.2% and 5.1%. This indicates that the VDVM effectively captures the axial loading characteristics across a broad envelope of advance ratios. While the VDVM consistently overpredicts $C_Q$ , the percentage difference remains stable, typically ranging from 11.5% to 14.13%. This systematic offset in torque suggests that while the model accurately tracks the aerodynamic trends, it may slightly struggle with secondary drag effects or skin friction components inherent in the CFD model. Overall, the strong correlation in thrust and the predictable nature of the torque deviations confirm the reliability of the VDVM for rapid aerodynamic assessment, with the minor discrepancies highlighting potential areas for tuning the profile drag coefficients.

Table 3: Comparison VDVM with CFD results

| $n$ | $J$ | $C_T$ (CFD) | $C_T$ (VDVM) | $C_Q$ (CFD) | $C_Q$ (VDVM) | $\Delta C_T\%$ | $\Delta C_Q\%$ |
|---|---|---|---|---|---|---|---|
| 3200 | 0.0000 | 0.03946 | 0.04148 | 0.00355 | 0.00396 | 0.0511 | 0.1154 |
| 2450 | 0.1506 | 0.03732 | 0.03897 | 0.00303 | 0.00340 | 0.0442 | 0.1221 |
| 2450 | 0.3011 | 0.03215 | 0.03330 | 0.00302 | 0.00344 | 0.0357 | 0.1390 |
| 2450 | 0.4517 | 0.02617 | 0.02713 | 0.00290 | 0.00331 | 0.0366 | 0.1413 |
| 2450 | 0.5576 | 0.02158 | 0.02227 | 0.00269 | 0.00303 | 0.0319 | 0.1263 |
| 2250 | 0.6072 | 0.01914 | 0.01983 | 0.00251 | 0.00283 | 0.0360 | 0.1274 |
| 1700 | 0.8037 | 0.00865 | 0.00909 | 0.00152 | 0.00172 | 0.0508 | 0.1315 |
| 1500 | 0.9108 | 0.00233 | 0.00244 | 0.00076 | 0.00086 | 0.0472 | 0.1315 |

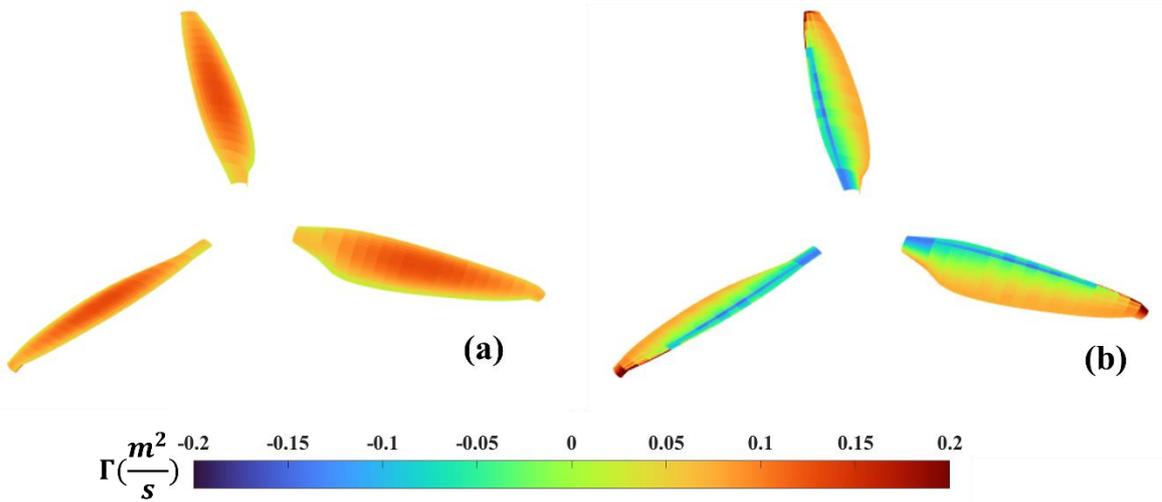

Fig. 3. Computed spanwise circulation distribution for (a) static conditions ($J= 0$, $n =3200$ rpm) and (b) forward flight ($J= 0.1506$, $n =2450$ rpm). The transition to a negative circulation zone in (b) illustrates the reduction in local effective angle of attack due to increased axial inflow.

The spanwise circulation distribution was evaluated for two distinct inflow conditions to validate the model's sensitivity to forward flight speeds. Figure 3(a) illustrates the circulation under static conditions ($J = 0$, $n = 3,200$ rpm), whereas Figure 3(b) displays the distribution at $J = 0.1506$ and $n = 2,450$ rpm. Under static conditions, the loading is predominantly positive across the span as the blade segments operate at high effective angles of attack. However, at $J = 0.1506$, a significant "negative zone" appears, particularly near the root and mid-span sections. This phenomenon occurs because the increased axial inflow velocity reduces the local effective angle of attack ($\alpha$) where the local geometric pitch angle ($\beta$) of the inboard sections is lower than the inflow angle, resulting in a negative angle of attack. Consequently, these sections generate negative lift (downforce) and negative circulation, effectively acting as a brake rather than a thrust producer. This transition underscores the necessity of including axial and tangential induction factors in the VDVM to accurately capture the shift from thrust-producing to drag-inducing regimes at high advance ratios.

## 5. Optimising Spanwise Chord and Twist Distributions

A systematic investigation of spanwise twist and chord distribution was conducted to quantify their effects on thrust and power. A parametric study, followed by optimization, was then used to identify design configurations that maximize thrust and propulsive efficiency.

### 5.1 Blade Twist Distribution Calculation

The twist distribution, $\beta$, is calculated by summing the two geometric contributions to the local pitch: the angle of attack ($\alpha$) and the resultant inflow angle ($\phi$):

$$\beta = \alpha + \phi \qquad (16)$$

In Eq. (16), $\phi$ is the resultant inflow angle at the blade section, and $\alpha$ is the local angle of attack. While $\alpha$ is determined at each section from the lift coefficient ($C_l$) distribution and the airfoil lift curve, $\phi$ is the remaining variable required to compute β. This angle is obtained from:

$$\phi = \arctan\left(\frac{U_\infty (1 + a)}{\omega r(1 - a')}\right) \qquad (17)$$

where $U_\infty$ is the free-stream velocity normal to the rotor plane (m·s⁻¹), $\omega$ is the angular speed (rad·s⁻¹), $r$ is the radial coordinate (m), and $a$ and $a'$ are the axial and tangential induction factors that quantify the propeller-induced velocity changes.

The axial induction factor $a$ accounts for the increase in axial flow velocity through the propeller disk, while the tangential induction factor $a'$ accounts for the swirl imparted to the flow in the tangential direction. These factors are typically determined iteratively using blade element momentum theory (BEMT) [28], where the momentum changes are balanced with the aerodynamic forces on the blade elements. The inflow angle $\phi$ is therefore evaluated at every spanwise station $r$ using Eq. (17), which properly accounts for both the axial and rotational induction effects on the local flow field. The resulting spanwise twist distribution, reflecting the optimized twist distribution ($\beta$) across the blade for maximum efficiency $(\eta)_{max}$, is presented in Fig. 4. This formulation is essential for accurate propeller design, as simplified models often neglect the significant induced velocities that occur under loaded conditions. In the baseline case, axial and tangential induction factors are not considered.

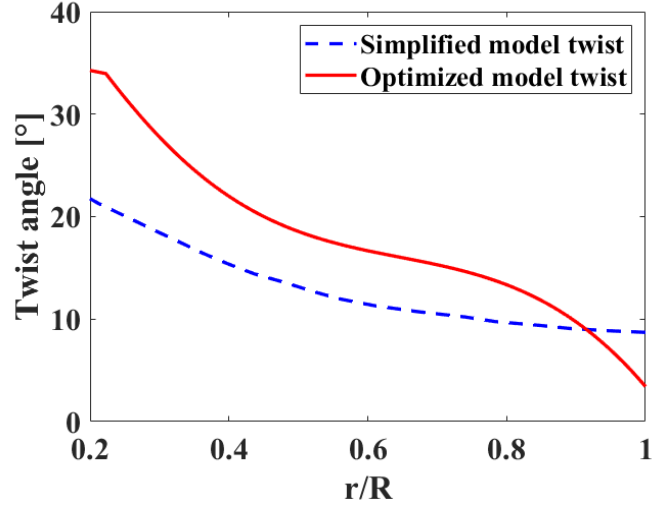

Fig.4. Spanwise twist distribution ($\beta$) incorporating axial and tangential induction effects across the propeller radius for $U_\infty = 1.5 \, m/s$ and $n = 1500$ rpm.

### 5.2 Blade Chord Distribution Calculation

The blade chord distribution is calculated by equating the required local circulation to the section lift production, ensuring the propeller satisfies the Betz condition [34] for maximum efficiency. Utilizing the iterative framework from the Adkins and Liebeck [28], the chord (c) at each radial station is determined using the fundamental relation:

$$c = \frac{4\pi \lambda G V R}{C_l \zeta B W} \qquad [18]$$

In this formulation, $\lambda$, represents the speed ratio ($U_\infty/\omega r$) and, $G$ represents the momentum loss factor derived from Prandtl [34] and Goldstein functions [35], $\zeta$ is the constant displacement velocity ratio across the wake, and $W$ is the local resultant velocity. This approach moves beyond light-loading approximations by explicitly accounting for induction effects and the local flow angle ($\phi$) through iterative computation [36]. By balancing the desired thrust or power requirements with the sectional lift coefficient $C_l$, the method produces a tapered chord distribution that decreases toward the blade tip, effectively optimizing the spanwise loading to minimize induced and profile drag losses. The optimized spanwise chord distribution and the baseline model [37] are illustrated in Fig. 5 for the prescribed loading conditions.

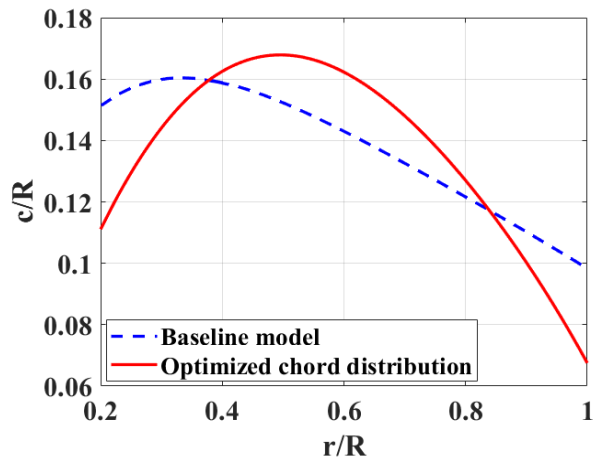

Fig.5. Baseline model [37] versus the Adkins-Liebeck derived chord profile, accounting for tip losses and induction effects.

## 6. RESULTS AND DISCUSSION

The performance of the Viscous Discrete Vortex Method (VDVM) was evaluated through a comparative analysis between the baseline rotor geometry and an optimized configuration. The primary objective was to assess how the integration of triple-deck viscous corrections influences the prediction of aerodynamic loads and how the subsequent shape optimization affects the overall propulsion efficiency.

### 6.1 Comparative Performance: Baseline vs. Optimized Model

The optimization process, focusing on the refinement of spanwise twist and chord distributions, yielded a configuration that significantly enhances aerodynamic efficiency. As summarized in Table 4, both models were evaluated at an advance ratios of 0.024.

Table 4: Performance Comparison between Baseline model with simplified twist and optimized model

| Configuration | $C_T$ | $C_Q$ | $C_P$ | $\eta$ |
|---|---|---|---|---|
| Baseline model with simplified twist | 0.14196 | 0.01143 | 0.07184 | 0.0473 |
| Optimized model | 0.14107 | 0.01043 | 0.06550 | 0.0515 |
| % Change | −0.63% | −8.75% | −8.82% | +8.99% |

The results indicate that while the $C_T$ remained nearly identical (a marginal decrease of 0.625%), the $C_P$ was reduced by 8.82 %. This leads to a substantial 9% increase in the efficiency ($\eta$). This improvement suggests that the optimized chord tapering and twist distribution effectively managed the spanwise loading, reducing induced losses and viscous drag without sacrificing the necessary thrust force.

### 6.2 Sensitivity to Rotational Speed

To evaluate the robustness of the VDVM and the performance of the optimized surfaces across varying operational regimes, the model was tested across an RPM range of 1610 to 2150, summarized in Table 5.

Table 5: Aerodynamic Loads across RPM Range (Optimized Geometry)

| $n$ | $J$ | $C_T$ | $C_Q$ | $C_P$ | $\eta$ |
|---|---|---|---|---|---|
| 1610 | 0.02291 | 0.14132 | 0.01044 | 0.06560 | 0.0494 |
| 1900 | 0.01941 | 0.14235 | 0.01063 | 0.06678 | 0.0414 |
| 1950 | 0.01892 | 0.14221 | 0.01066 | 0.06699 | 0.0402 |

| $n$ | $J$ | $C_T$ | $C_Q$ | $C_P$ | $\eta$ |
|---|---|---|---|---|---|
| 2000 | 0.01844 | 0.14186 | 0.01071 | 0.06731 | 0.0389 |
| 2050 | 0.01799 | 0.14092 | 0.01077 | 0.06765 | 0.0375 |
| 2150 | 0.01716 | 0.13888 | 0.01097 | 0.06895 | 0.0346 |

The relationship between rotational speed and aerodynamic loads follows the expected non-linear trend. As the $n$ increases from 1610 to 2150 rpm (a 33% increase), the $C_T$ reduced by approximately 1.78 % while the $C_P$ increased by 5.10 %. This highlights the critical importance of optimization at higher speeds, where power penalties for inefficient designs grow exponentially.

### 6.3 Aerodynamic Trade-offs and Geometric Insights

The parametric investigation revealed that the viscous-inviscid interaction at the trailing edge, captured by the triple-deck correction, plays a vital role in determining the aerodynamic performance limits and the onset of flow separation.

- Chord Distribution: The transition from the baseline to the tapered chord distribution (calculated via Eq. 18) successfully reduced tip losses. By incorporating a tapered chord toward the tip, the VDVM captured a more elliptical thrust distribution, which is theoretically optimal for minimizing induced drag.

- Twist Distribution: The twist distribution was determined using Eq. 16, which explicitly incorporates both axial ($a$) and tangential ($a'$) induction factors. This ensured that each blade section operated at its optimal angle of attack relative to the local resultant velocity, preventing premature local separation that inviscid models often fail to predict.

By replacing the classical Kutta condition with the viscous closure, the VDVM provided a more realistic estimation of the pressure jump near the trailing edge, particularly at the higher RPM cases where the effective angle of attack is significant. This correction prevented the overprediction of thrust commonly seen in inviscid discrete vortex methods.

### 6.4 Spanwise Loading Distribution

The spanwise distribution of aerodynamic loads for the optimized rotor configuration is illustrated in Fig. 6a and 6b, representing the sectional thrust and torque profiles calculated using the VDVM. As illustrated in Fig. 6a, the sectional thrust distribution exhibits a non-linear profile; loading remains minimal near the hub before reaching a peak of approximately 145 N within the 0.75 to 0.85 r/R radial span. This peak reflects the effective integration of the calculated twist distribution and induction factors, which maximize local angle of attack efficiency in high-velocity regions. A significant reduction in sectional thrust is observed beyond the 0.85 r/R radial span, reaching its minimum at the blade tip, a direct result of the tapered chord distribution derived from the Adkins and Liebeck framework [28] used to mitigate tip-vortex induced losses. Similarly, the torque distribution shown in Fig. 6b peaks at approximately 14 Nm near the 1.0 m station. This demonstrates the aerodynamic cost of thrust generation, as influenced by the local resultant velocity and the viscous drag predicted by the triple-deck correction. The significant reduction in torque toward the tip confirms that the optimized chord tapering successfully reduces profile and induced drag penalties, directly contributing to the 8.99% improvement in the efficiency observed in the final optimized model.

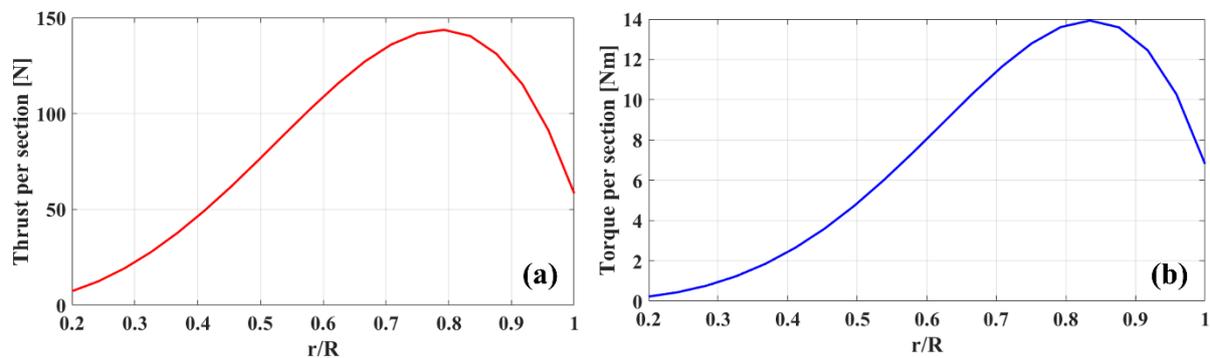

Fig.6 Spanwise distribution of (a) thrust and (b) torque for the optimized rotor geometry, calculated using the Viscous Discrete Vortex Method (VDVM).

### 6.5 Spanwise Circulation Distribution

The spanwise circulation distribution (Γ) over the optimized rotor blade is illustrated in the figure.7, providing critical insight into the load distribution. The normalized circulation values range from approximately -0.4 to 0.4. The concentrated red areas typically correspond to the mid-to-outboard sections of the blade where the combination of local tangential velocity and optimized chord length results in peak bound circulation. This region represents the primary lifting zone of the rotor, where the pressure jump across the blade is maximized. The blue regions near the root or trailing edges represent areas of significantly lower or localized negative circulation. This distribution is a direct result of the Viscous Discrete Vortex Method (VDVM), which accounts for viscous-inviscid interactions at the trailing edge by relaxing the classical Kutta condition. The transition from red to blue toward the blade tip highlights the effectiveness of the tapered chord distribution derived from the Adkins and Liebeck framework [28]. By forcing the circulation to gradually diminish toward the tip, the model effectively captures the mitigation of tip-vortex strength, which contributed to the 8.99% improvement in $\eta$ observed in this study. This smooth gradient ensures a more efficient, nearly elliptical thrust distribution that minimizes induced drag across the span.

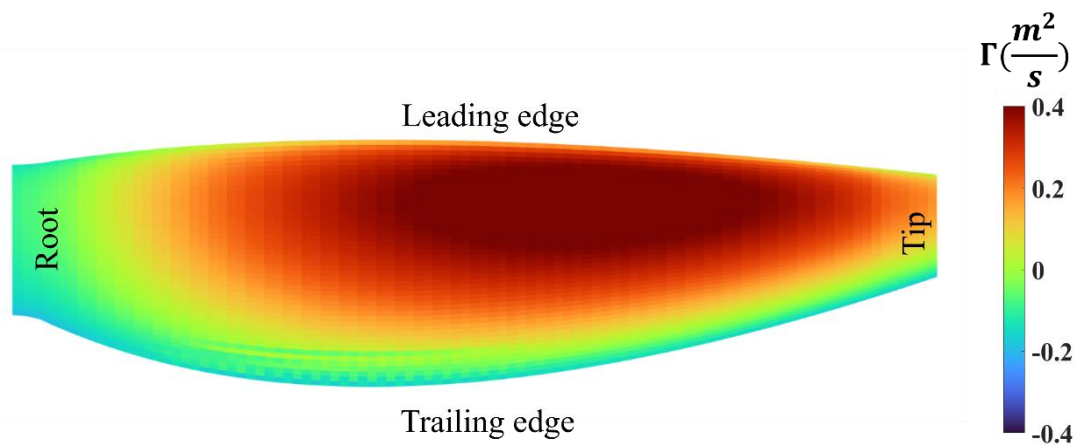

Fig.7. Aerodynamic loading visualization illustrating the spanwise circulation distribution over the optimized blade.

## 7. Conclusion

This study successfully developed and validated a Viscous Extension Discrete Vortex Method (VDVM), bridging the gap between computationally expensive high-fidelity CFD and simplified inviscid potential flow models. By replacing the classical Kutta condition with a viscous closure derived from triple-deck boundary layer theory, the model introduces a critical dependence on the Reynolds number and effective angles of attack.

The primary findings of this research are summarized as follows:

- The VDVM shows good accuracy across a broad operational range (1330–8841 RPM). Experimental validation yielded $C_T$ deviations of only 1.99% to 4.3%, while $C_Q$ remained within 7.95% to 14.8% due to viscous modeling sensitivities.

- Validation against CFD data was conducted for advance ratio s ranging from 0 to 0.9108. For $C_T$, results showed strong agreement at rotational speeds between 1500 and 3200 RPM, with error margins of 3.2% and 5.2%, respectively. While the VDVM consistently overpredicts $C_Q$, the percentage difference remains stable, typically ranging from 11.6% to 14.1%.

- The optimization of spanwise twist and chord distributions, accounting for both axial and tangential induction factors, resulted in a significant performance enhancement. The optimized rotor achieved an 8.99% increase in the efficiency ($\eta$) compared to the baseline model.

- The implementation of a tapered chord distribution (via the Adkins and Liebeck framework) and a non-linear twist profile proved essential in managing spanwise loading. These modifications minimized induced losses at the blade tips and balanced the sectional thrust coefficient to prevent premature flow separation.

- VDVM provides a robust and fast-running tool for the iterative design phase of rotor blades. It captures the non-linear aerodynamic effects of high-speed rotation without the prohibitive time costs of Navier-Stokes simulations.

We believe this integrated viscous-inviscid framework provides an essential tool for the rapid development of high-efficiency eVTOL propellers, where the ability to balance computational speed with viscous accuracy is critical for exploring the vast design spaces of distributed electric propulsion.